\documentclass[conference]{IEEEtran}
\ifCLASSINFOpdf
\else
\fi
\usepackage{graphicx}

\newcommand{\RomanNumeralCaps}[1]{\MakeUppercase{\romannumeral #1}}
\usepackage{multirow}
\usepackage{array}
\usepackage{tabularx}

\begin{document}
%
\title{How mobility patterns drive disease spread: A case study using public transit passenger card travel data}

\author{
    \IEEEauthorblockN{Ahmad El Shoghri\IEEEauthorrefmark{1}\IEEEauthorrefmark{2}, Jessica Liebig\IEEEauthorrefmark{2}, Lauren Gardner\IEEEauthorrefmark{3}\IEEEauthorrefmark{4}, Raja Jurdak\IEEEauthorrefmark{2}, Salil Kanhere\IEEEauthorrefmark{1}}
    \IEEEauthorblockA{\IEEEauthorrefmark{1}School of Computer Science and Engineering, University of New South Wales, Sydney, NSW, AUSTRALIA
    \\ahmad.elshoghri@student.unsw.edu.au}
    \IEEEauthorblockA{\IEEEauthorrefmark{2}Data61, Commonwealth Scientific and Industrial Research Organization, Brisbane, QLD, AUSTRALIA
    }
    \IEEEauthorblockA{\IEEEauthorrefmark{3}School of Civil and Environmental Engineering, University of New South Wales, Sydney, NSW, Australia
    }
    \IEEEauthorblockA{\IEEEauthorrefmark{4}Department of Civil Engineering, Johns Hopkins University, Baltimore, MD, USA
    }
}

\IEEEoverridecommandlockouts
\IEEEpubid{\makebox[\columnwidth]{978-1-7281-0270-2/19/\$31.00 \copyright2019 IEEE \hfill} 
\hspace{\columnsep}\makebox[\columnwidth]{ }}

\maketitle
\IEEEpubidadjcol
\begin{abstract}
Outbreaks of infectious diseases present a global threat to human health and are considered a major health-care challenge. One major driver for the rapid spatial spread of diseases is human mobility. In particular, the travel patterns of individuals determine their spreading potential to a great extent. These travel behaviors can be captured and modelled using novel location-based data sources, e.g., smart travel cards, social media, etc. Previous studies have shown that individuals who cannot be characterized by their most frequently visited locations spread diseases farther and faster; however, these studies are based on GPS data and mobile call records which have position uncertainty and do not capture explicit contacts. It is unclear if the same conclusions hold for large scale real-world transport networks. In this paper, we investigate how mobility patterns impact disease spread in a large-scale public transit network of empirical data traces. In contrast to previous findings, our results reveal that individuals with mobility patterns characterized by their most frequently visited locations and who typically travel large distances pose the highest spreading risk.
\end{abstract}


%
\IEEEpeerreviewmaketitle

\section{Introduction}
Infectious diseases such as influenza, measles and tuberculosis pose an ongoing threat to people and global health security \cite{gog2014spatial}\cite{grenfell2001travelling}\cite{prothero1977disease}. Studies show that epidemic outbreaks in many countries are considered a major health-care challenge that causes morbidity and mortality. Existing methods used to identify the main factors that lead to disease outbreaks are showing vulnerabilities \cite{heesterbeek2015modeling} and various outbreaks were witnessed in recent years \cite{wesolowski2016connecting}.


The spread of an infectious disease is highly opportunistic and heterogeneous \cite{rubrichi2017comparison}. While several factors shape the spreading dynamics of infectious diseases, human movement remains the key driver behind the prevalence of any infectious disease \cite{wesolowski2016connecting}. The spreading dynamics of a disease can be affected in different ways, for example, by the change of contact frequency among infected and susceptible individuals or through the introduction of new pathogens into susceptible regions \cite{wesolowski2016connecting}\cite{rubrichi2017comparison}\cite{bansal2016big}. 
The growing popularity of location-based applications, through which human movement traces are available, led to a rapid emergence of big data in science research in recent years \cite{bansal2016big}\cite{ghosh2017modeling}\cite{thilakarathna2017deep}. This data provides an unprecedented opportunity to derive spatial and temporal knowledge that can be related directly to the risk of disease spread \cite{wesolowski2016connecting}. Specifically, elucidating the spatial travel patterns of individuals will reveal the travel behaviors that pose the highest risk in a disease spread scenario \cite{wesolowski2016connecting}\cite{bansal2016big}. This information is beneficial to the concerned authorities to potentially forecast the risk of a disease outbreak and to develop targeted prevention and containment strategies \cite{rubrichi2017comparison}\cite{wesolowski2014commentary}.

The current state-of-the-art suggests that individuals whose mobility patterns cannot be characterized by their most frequently visited locations, or whose recurrent mobility does not dominate their total mobility, have the highest disease spreading ability \cite{pappalardo2015returners}. However, the interaction of this behavior with other important spatial features such as the traveled distances is unknown. In this paper, we use empirical large-scale human movement data to perform trace driven simulations for spreading a disease on a public transport network. This study contributes to the understanding of how different mobility patterns of individuals influence the spreading dynamics of a disease. Our results are obtained through extensive simulations using one month of comprehensive bus passenger data collected in Sydney, Australia. From this data, we exploit the co-presence of passengers traveling on the same bus to construct a city-scale time-resolved physical human contact network on which we simulate the spread of a disease.

In the remainder of the paper we explore the effect of the individuals' traveled distances on the spreading dynamics. We then investigate the behavior of two distinct mobility patterns called returners and explorers along different traveled distances. Finally, we reveal the type of passenger movement that has the highest impact on spreading a disease in a public transport network.

 

\section{Related Work}

Several studies have recognized that human mobility highly affects the spread of infectious diseases \cite{gog2014spatial}\cite{wesolowski2016connecting}\cite{mari2011modelling}\cite{prothero1977disease}\cite{grenfell2001travelling}\cite{stoddard2009role} \cite{wesolowski2014commentary}. However, obtaining data that accurately captures human mobility traces and physical encounters between individuals remains a challenge due to privacy and confidentiality regulations \cite{tizzoni2014use}. Previous studies have tried to reconstruct travel patterns of individuals as well as human contact networks from mobile phone calls, global positioning system (GPS) data and the circulation of bank notes \cite{gonzalez2008understanding}\cite{brockmann2006scaling}.

An important body of research has explored the use of call detail records (CDRs) to build epidemiological models and to study the spatial transmissions of various spreading diseases in a population at both city and country levels \cite{wesolowski2012quantifying}\cite{tatem2014integrating}\cite{wesolowski2015impact}\cite{bengtsson2015using}\cite{wesolowski2015quantifying}\cite{isdory2015impact}\cite{brdar2016unveiling}. 
The authors of \cite{pappalardo2015returners} found two distinct human mobility behaviors within CDRs and GPS datasets that were both collected in Italy. An individual is either a returner, that is his recurrent mobility dominates his total mobility, or an explorer, in which case his recurrent mobility does not dominate his total mobility. The statistical measure used to calculate the mobility of individuals is the total radius of gyration, which is the measure of deviation of an individual from the center of mass of his visited locations \cite{gonzalez2008understanding}. The total radius of gyration $r_g$ is defined as \cite{pappalardo2015returners}:\\

\begin{equation}
r_g = \sqrt{ \frac{1}{N} \sum_{i \in L}{ n_i (r_i - r_{cm})^2 } }
\end{equation}
where $r_i$ are the coordinates of the visited locations, $n_i$ are the respective visitation frequencies, $r_{cm}$ is the center of mass of the set of visited locations and $N$ is the total number of visited locations. This measure characterizes the overall mobility of each individual. To understand the effect of the frequently visited locations on the mobility range of the individuals, the authors of \cite{pappalardo2015returners} define the $k$-radius of gyration, ${r_g}^{(k)}$, where the computation of the radius of gyration is limited to the $k$ most frequently visited locations. The $k$-radius of gyration characterizes the recurrent mobility of each individual. When the mobility values are plotted in two-dimensional space, i.e. recurrent mobility on the $y$-axis and total mobility on the $x$-axis, one can distinguish between two distinct mobility behaviors, namely returners and explorers. The data points concentrated along the $x$-axis correspond to individuals whose $k$-radius of gyration is smaller than their total radius of gyration. In other words, their mobility patterns cannot be characterized by their most frequently visited locations. These individuals are called explorers \cite{pappalardo2015returners}. The data points concentrated around the $y=x$ line correspond to individuals whose characteristic traveled distance is dominated by their most frequently visited locations, ${r_g}^{(k)} \approx r_g$. The mobility behavior of these individuals can be reduced to their $k$ most frequently visited locations. Such individuals are called returners \cite{pappalardo2015returners}. The work suggests using a bisector method to classify each individual to the corresponding mobility behavior.
While the authors of \cite{pappalardo2015returners} found that explorers are more likely to spread a potential disease on the contact networks that were reconstructed from the CDRs and GPS data, it is unclear if this finding is also true for other, more realistic contact networks. 
There are several limitations in the way the networks are constructed, mainly cellular towers are often located hundreds of meters apart from each other which makes it impossible to accurately locate individuals and determine if two individuals were in close enough proximity to transmit a disease [21]. 
Similarly, analyzing human mobility patterns and spreading dynamics using datasets collected from vehicles equipped with GPS presents various limitations. First, this type of data is vulnerable to spatial errors such as inaccurate localization due to poor satellite signals and missing data due to complete loss of signals \cite{vazquez2013using}. Second, since GPS devices are placed on vehicles, the data does not capture the number of individuals embarking the same vehicle and does not guarantee the occurrence of a physical human contact between individuals embarking two different vehicles \cite{tizzoni2014use}. These limitations make the GPS data ineffective when studying the spreading dynamics of a contagious disease \cite{tizzoni2014use}. 

Recent studies of human mobility and disease spread demonstrated an increasing interest in real physical encounters. In contrast to non-physical contacts initiated by mobile phone calls, e-mails, and online social networks etc., physical contact networks do not assume previous friendship or familiarity between two individuals however their encounter is accurately detected upon their co-presence in both spatial and temporal dimensions \cite{sun2014efficient}. Therefore, additional studies based on large-scale empirical data that captures real physical human encounters are still needed \cite{wesolowski2016connecting}. Our large scale bus passenger data is well suited to fill this gap as it records real physical human contacts between individuals.
\section{Returners and explorers in new dimensions}
\subsection{Public Transit Traveler Data}
In this study we use public transit smart card data that includes tap-on and tap-off timestamps for all bus passengers traveling in the greater metropolitan region of Sydney, Australia during the month of April in 2017. Tap-on and tap-off timestamps refer to the times at which a passenger enters and exits a bus respectively. This dataset consists of 20,295,908 trips belonging to 2,010,541 users. In particular, each trip record contains the following information: passenger ID, tap-on time, tap-on location, tap-off time, tap-off location and vehicle ID. The high spatio-temporal resolution of this dataset allows us to extract time-resolved encounters in buses, defined as two individuals occupying the same bus simultaneously. Using this information we create a city-scale contact network on which we simulate the spread of a disease.
\subsection{Simulation Setup - network, timeline and disease spread model}
Our study focuses on simulating the spread of a disease on a physical contact network constructed from real human movement data traces. For this reason, we modified the opportunistic network environment (ONE) simulator targeted for research in delay tolerant networks to support trace driven simulations for spreading a disease on a public transport network. Each experiment in our study is simulated 100 times and the results are averaged. The simulations start by selecting a population. In our experiments, we include the contact links that are connecting the set of individuals chosen. At the start of each simulation 500 seed nodes are randomly chosen to be infected with a disease that will propagate through the network following a Susceptible-Infectious-Recovered (S-I-R) simulation model to represent disease spread in a population of individuals. Following the S-I-R model every individual is in one of three possible states, susceptible (S), infected (I) or recovered (R). In our simulation setup, if a susceptible and an infected individual meet, the infection is transmitted immediately and the susceptible individual changes its state to being infected. Once infected, the individual remains infectious for five days before changing to the recovered state. The individual remains recovered until the end of the simulation. Since we are not interested in a particular disease and to simplify the comparison between travel behaviors we choose a transmission probability of one, which represents the maximum spreading case scenario.
\subsection{Identifying Returners and Explorers}
\begin{figure}
\centering
\includegraphics[width=\linewidth]{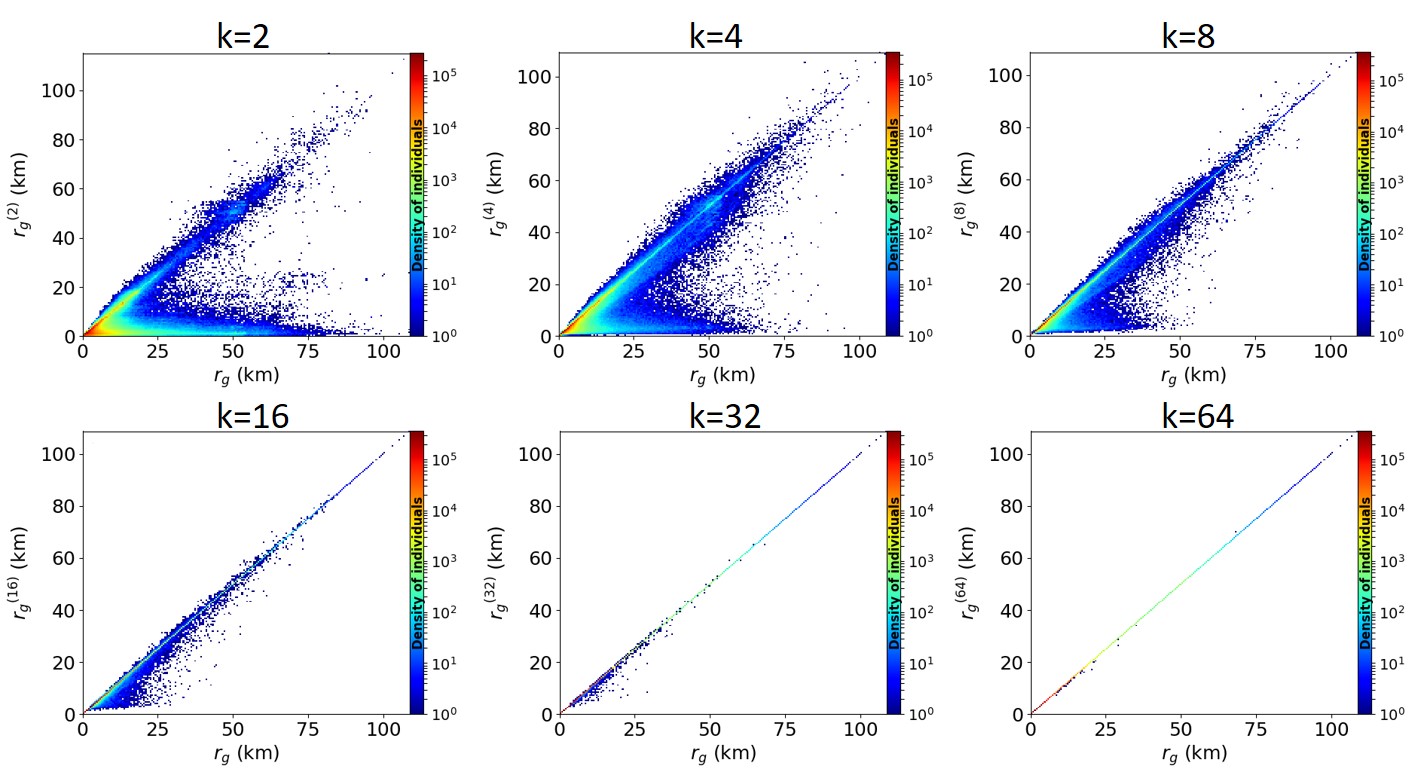}
\caption{The correlation between recurrent mobility (${r_g}^{(k)}$) and overall mobility ($r_g$) for $k$=2, 4, 8, 16, 32 and 64. The colors range from red to blue to indicate the density of the points in the corresponding area, with red referring to high density. The points along the $x$-axis correspond to explorers and the points along the $y=x$ line correspond to returners.}
\label{fig:opalExpRetCorr}
\end{figure}

\begin{figure}
\centering
\includegraphics[width=\linewidth]{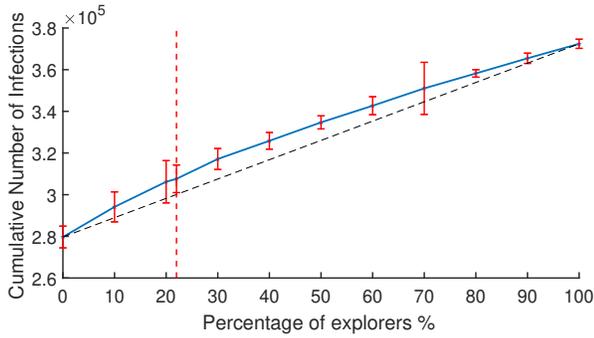}
\caption{The cumulative number of infections for varying percentage of explorers. The dotted red line represents the actual percentage of explorers in the full travel card dataset, 22\% of the total population. The full red lines represent the error bars, reporting the uncertainty in the cumulative number of infections for each percentage of explorers.}
\label{fig:opalCumInfExpRet}
\end{figure}

From previous studies it is unclear to what extent returner and explorer mobility patterns observed in CDRs and GPS data extend to transport networks. For this reason, we first compute the values of the total radius of gyration and various $k$-radii of gyration for each bus passenger in the smart travel card data.
Figure \ref{fig:opalExpRetCorr} shows the correlation between the total radius of gyration and $k$-radius of gyration of each individual in the bus transportation network for $k$ = 2, 4, 8, 16, 32 and 64. The plots in Fig. \ref{fig:opalExpRetCorr} clearly confirm the existence of returners (points concentrated along $y=x$ line) and explorers (points concentrated along the $x$-axis) in the public bus transportation network of Sydney. As the value of $k$ increases the individuals move from being explorers to being returners since their recurrent mobility over a larger number of visited locations is closer to represent their overall mobility. To study the role of each of the two groups in a realistic disease spreading environment, we create a contact based disease transmission scenario in which people travel on a bus network according to real world bus activity traces.
We run several simulations in this experiment varying the percentage of explorers. We start with 0\% explorers and 100\% returners, chosen randomly from the full set of passengers classified as returners in the original network. We then increase the percentage of explorers by 10\% at a time, until explorers constitute the whole population. In each simulation the remaining percentage of the population is strictly chosen from the returners until the total number of individuals is reached. We choose $k$=2 for this experiment. The population consists of 400,000 individuals due to the limit in the number of travelers classified as explorers in the data. The disease propagates through the network as described in \mbox{Section \RomanNumeralCaps{3}. $B$}. Due to the randomness resulting from the chosen passenger set, we average the results over 100 simulations.
Fig. \ref{fig:opalCumInfExpRet} shows that the cumulative number of infections increases with the increase of explorers in the network, highlighting the distinct role that each of the two types of mobility behaviors plays in spreading a disease in the transport network with explorers being more influential. The dotted red line represents the actual percentage of explorers in the full smart travel card dataset which is 22\% of the total population. 

Sub-sampling a population may involve isolated nodes in the network. To explore this aspect we introduce in \mbox{Table \ref{tbl:excel-table}} the number of isolated travelers who are a result of including only contact links connecting the randomly chosen individuals. The percentage of isolated individuals range from 0.1\% at 100\% explorers to 3\% at 0\% explorers. Figure \ref{fig:opalCumInfExpRet} shows that if the network consists of 100\% explorers the cumulative number of infections reaches approximately 370,000. That is, 92.58\% of non-isolated individuals are infected. In contrast, if the network consists of 100\% returners (i.e. 0\% explorers) the cumulative number reaches approximately 280,000 infections, which is 72.4\% of non-isolated individuals. The spreading percentages show that the networks are well connected although some isolated nodes (less than 3\%) exist. Further investigation shows similar degree distributions for the different sub-networks: many nodes have few contact links and relatively few nodes have many contact links. These observations confirm that explorers pose a higher disease spreading risk and that the results are not an artefact of population selections.
\begin{table}
  \caption{The number of isolated individuals in the population.}
  \label{tbl:excel-table}
  \includegraphics[width=\linewidth]{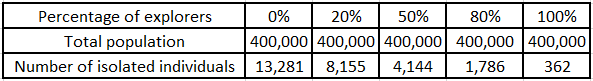}
\end{table}

\subsection{Distance, total radius of gyration}
\begin{figure}
\centering
\includegraphics[width=8cm,height=4cm]{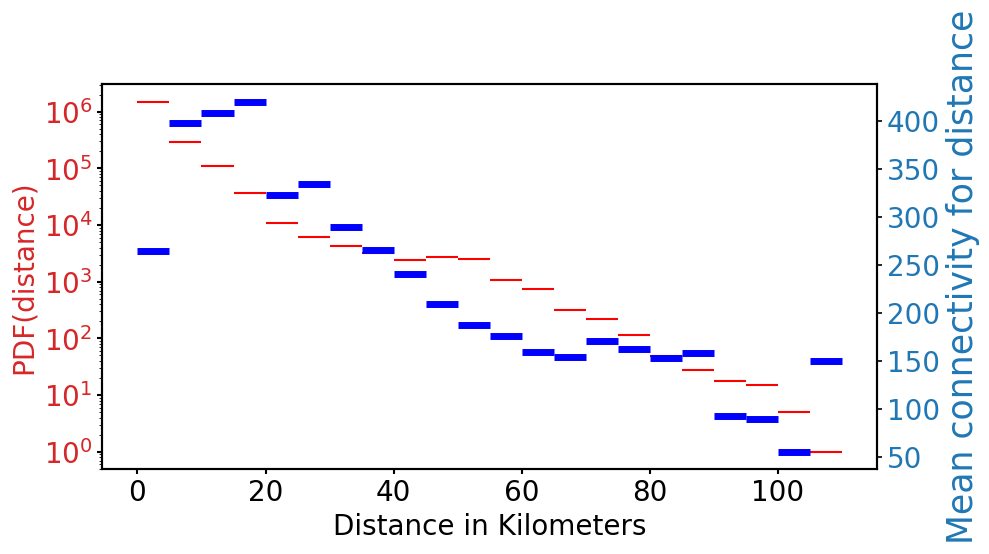}
\caption{The distribution of distances and the mean contact links in terms of distance.}
\label{fig:opalRgConnDist}
\end{figure}

\begin{figure}
\centering
\includegraphics[width=\linewidth]{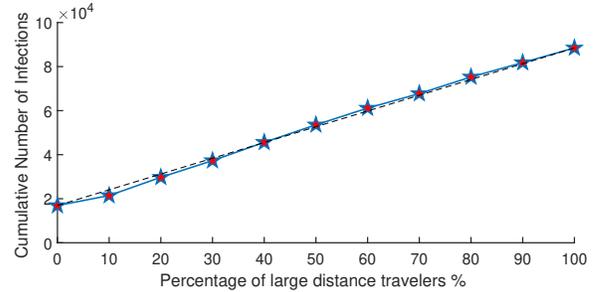}
\caption{The cumulative number of infections for varying percentages of large distance travelers.}
\label{fig:opalSmallLargeDistCumInf}
\end{figure}

In this section, we expand our analysis to the distance dimension. In particular, we highlight the spreading behavior as a function of the radius of gyration. To gain insight into the distribution of the radius of gyration among the individuals, we group the values in bins of 5 KM and plot the probability density function (pdf) of the total radius of gyration against the mean number of encounters for each group. The plots in Fig. \ref{fig:opalRgConnDist} indicate that most of the individuals in the bus dataset move within short distances and relatively fewer individuals tend to travel long distances (and cover larger areas). Furthermore, the individuals who cover short distances have more encounter opportunities in comparison to other individuals who travel longer distances due to moving more centrally. To be able to compare the effect of distance alone on the spreading dynamics, we include only those individuals who encountered 10 to 400 other individuals. This range of encounters is the only range that covers the full distance spectrum, i.e., small and large distances. We apply a standard $K$-means clustering algorithm \cite{pedregosa2011scikit} on the total radius of gyration of the selected individuals, with $K$=2, which splits the population into two groups: individuals radiating in small distances and those who radiate in large distances.
The value of separation between small and large distances is approximately 12 KM. To investigate the effect of each travel pattern, namely small distance and large distance travelers, we conduct a sensitivity analysis to elucidate the role of passenger travel distance on infectious disease spreading risk. We set the population size to 124,757 individuals (limited by the number of large distance individuals) and vary the percentage of large distance individuals selected in the simulation. A percentage of 0\% means that the whole population is chosen from the small distances group and 100\% means that the whole population consists of individuals from the large distances group. Next, we compare the values of the cumulative number of infections at the end of the simulation. The results in Fig. \ref{fig:opalSmallLargeDistCumInf} clearly show that large distance travelers increase the chances of spreading the infection. This is due to the existence of more individuals moving in large radii, allowing the infection to reach farther areas and to connect more sub-populations. This leads to new encounter opportunities and hence a higher cumulative number of infections. In contrast, the movement made by small distance travelers limits the propagation of the infection. This type of movement forces the infection to stay in local areas, circulate among recovered or already infected individuals and restricts it from spreading widely in the network.

\subsection{Distance dimension with returners and explorers}
\begin{figure}
\centering
\includegraphics[width=0.6\linewidth]{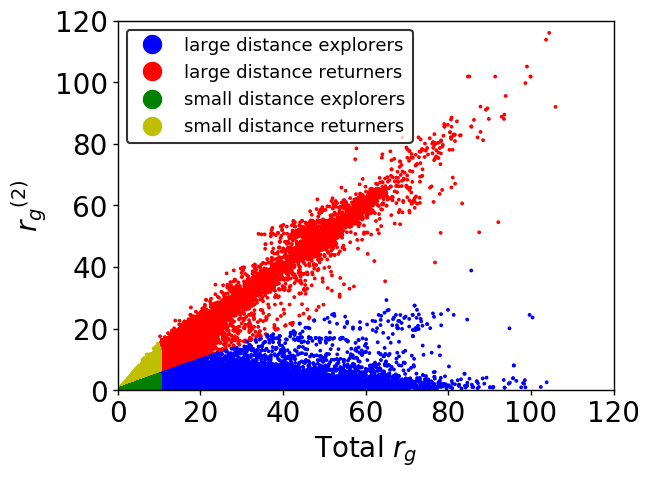}
\caption{The population split into four main groups: small distance explorers 15.54\% (green points), small distance returners 76.35\% (yellow points), large distance explorers 3.88\% (blue points) and large distance returners 4.21\% (red points).}
\label{fig:opalClustAll}
\end{figure}

In this section, we expand our study of the distance by considering a further split among the individuals based on the returners and explorers dichotomy. We differentiate between small distance explorers, small distance returners, large distance explorers and large distance returners. The aim of this experiment is to investigate in more detail the impact that each group has on the spread.
To derive the four mobility patterns, we plot the correlation between recurrent mobility and overall mobility of the individuals. Then, we apply the standard $K$-means algorithm \cite{pedregosa2011scikit} to split the individuals based on the spatial dimension. We then use the bisector method to further split the individuals between returners and explorers \cite{pappalardo2015returners}. The four groups that appear after this procedure with their respective percentages out of the total population are shown in Fig. \ref{fig:opalClustAll}.

\begin{figure}
\centering
\includegraphics[width=\linewidth]{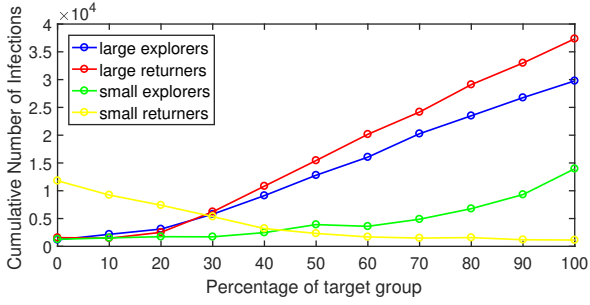}
\caption{The cumulative number of infections for varying percentages of small distance explorers (green line), small distance returners (yellow line), large distance explorers (blue line) and large distance returners (red line).}
\label{fig:opalAllCumInf}
\end{figure}

\begin{figure}
\centering
\includegraphics[width=\linewidth]{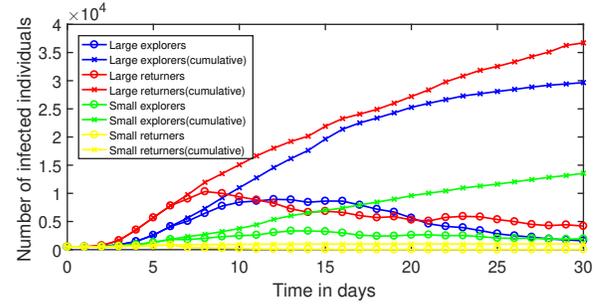}
\caption{The detailed spreading behaviors of small distance explorers (green line), small distance returners (yellow line), large distance explorers (blue line) and large distance returners (red line) at a target group percentage of 100\%. The crossed lines correspond to the cumulative number of infections since the start of the simulation and until time $t$ and the circled lines correspond to the actual number of all infectious individuals present in the simulation at time $t$.}
\label{fig:opalAllInf}
\end{figure}

In this experiment, the total population has 60,000 individuals (limited by the number of individuals classified as large distance explorers). This population number in each simulation consists of individuals from a target group and individuals from the remaining groups. Target group refers to the cohort for which we are varying the percentage to study its spreading behavior and to assess its impact on the scenario. The remaining individuals are chosen from the other three groups only until the total population number is reached. The remaining individuals are chosen randomly to ensure a mix of the three types of individuals and to preserve their initial fraction of the population. For example, the red circle at 50\% in Fig. \ref{fig:opalAllCumInf} summarizes the simulations in which the total population of 60,000 consists of 30,000 individuals from the large distance returners group and 30,000 individuals chosen randomly from the combination of the three remaining groups. The percentages of the target group vary from 0\% at which none of the individuals are chosen from that group to 100\% where the total population is chosen from the target group. Figure \ref{fig:opalAllCumInf} shows the average cumulative number of infections at the end of the simulations of the four groups. We notice that the increase in the percentage of each target group results in an increase in the cumulative number of infections for all groups except the small returners (yellow line). The cumulative number of infections of large returners reaches higher values through all the experiments, especially from 30\% and above. The decreasing trend in the cumulative number of infections for small returners indicates a negative impact on the spreading. The highest value of the number of infections was at 0\% at which no individuals from that group were chosen. The relatively higher values of infections at 40\% and below are therefore due to the effect of the other three groups on the spreading dynamics. The cumulative number of infections decreases as we include more small returners in the simulations. Figure \ref{fig:opalAllInf} presents the detailed spreading behaviors of the four different groups at a target group percentage of 100\%. The crossed lines correspond to the cumulative number of infections since the start of the simulation and until time $t$ and the circled lines correspond to the actual number of all infectious individuals present in the simulation at time $t$. We notice in Fig. \ref{fig:opalAllInf} that the crossed lines and the circled lines overlap up to 5 days at the start of the simulation. This observation is due to the 5 days infectious period during which all infected individuals stayed infectious and none was able to recover (see \mbox{Section \RomanNumeralCaps{3}. $B$}).

\begin{figure}
\centering
\includegraphics[width=8.2cm,height=4cm]{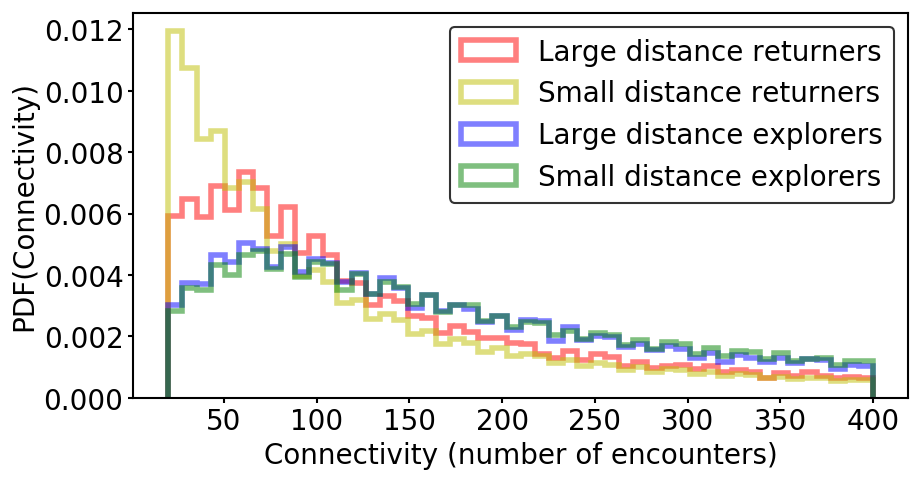}
\caption{The contact link distribution for small distance explorers (green line), small distance returners (yellow line), large distance explorers (blue line) and large distance returners (red line).}
\label{fig:opalAllConnDist}
\end{figure}
\begin{table}
  \centering
  \caption{The frequency of large distance trips.}
  \label{tbl:excel-table_freq}
  \includegraphics[width=0.8\linewidth]{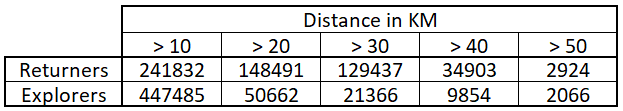}
\end{table}

Two key findings can be highlighted in this experiment. First, explorers are not the only main spreaders as was found in previous studies. In fact, our analysis shows that returners play the most important role but only those who move in large distances. This was not previously found due to the limitations that arise from using CDR and GPS datasets as well as unrealistic network assumptions (see \mbox{Section \RomanNumeralCaps{2}}). Second, small distance returners can be excluded from containment strategies since they have the least effect on the diffusion in a disease spread scenario. Our results improve the understanding of the key factors and features responsible for spreading a disease. Our findings can help in the design of efficient prevention and containment strategies that guarantee the best health safety outcomes.
To show that the results are not strictly due to having a higher number of encounters for large distance returners we plot in Fig. \ref{fig:opalAllConnDist} a step histogram showing the probability distribution function of the connectivity for each of the four groups until reaching 400 encounters which is the upper bound for our experiment. We notice that returners in general are less likely to encounter a large number of individuals as compared to explorers. Furthermore, we investigate the difference in the frequency of long distance trips made by returners and explorers. \mbox{Table \ref{tbl:excel-table_freq}} shows the frequencies of the trips made in terms of traveled distance in kilometers. The table shows higher frequency values for returners for trips having at least 10 KM distance. This result suggests that large distance returners are consistent in making regular long distance trips as compared to explorers who tend to make fewer long distance trips.
These observations confirm that the impact of large distance returners on spreading is due to the mobility behavior of that group and shed light on the unique transmission opportunities that are created through the high frequency of actual long distance trips as shown in \mbox{Table \ref{tbl:excel-table_freq}}.

\section{Conclusion}
In this work, we investigated large-scale human contact networks that were constructed from smart travel card data collected in Sydney, Australia. We categorized 2,010,541 individuals based on their mobility behavior and performed extensive disease spread simulations on the constructed contact networks. The results of this unprecedented study contribute to the understanding of how different mobility patterns of individuals influence the spreading dynamics of contagious diseases. In contrast to previous results, our study shows that individuals classified as returners who also cover long distances spread diseases farther and faster. 





%



\medskip

\bibliographystyle{unsrt}
\bibliography{mybibliography}

\end{document}